\documentclass[aps,prd,showpacs,nofootinbib,preprint]{revtex4}

\usepackage{amsmath}
\usepackage{amssymb}
\usepackage{graphicx}
\usepackage{verbatim}

\begin{document}
\pagestyle{plain}

\title{Inflation and late time acceleration in braneworld cosmological models with varying brane tension}
\author{K. C. Wong}
\email{fankywong@gmail.com}
\affiliation{Department of Physics, University
of Hong Kong, Pok Fu Lam Road, Hong Kong, P. R. China}
\author{K. S. Cheng}
\email{hrspksc@hkucc.hku.hk}
\affiliation{Department of Physics
and Center for Theoretical and Computational Physics, University
of Hong Kong, Pok Fu Lam Road, Hong Kong, P. R. China}
\author{T. Harko}
\email{harko@hkucc.hku.hk}
\affiliation{Department of Physics and
Center for Theoretical and Computational Physics, University of
Hong Kong, Pok Fu Lam Road, Hong Kong, P. R. China}
\date{\today}

\begin{abstract}
Braneworld models with variable brane tension $\lambda $ introduce a new degree of freedom that allows for evolving gravitational and cosmological constants, the latter being a natural candidate for dark energy.  We consider a thermodynamic interpretation of the varying brane tension models,  by showing that the field equations with variable $\lambda $ can be interpreted as describing matter creation in a cosmological framework. The particle creation rate is determined by the variation rate of the brane tension, as well as by the brane-bulk energy-matter transfer rate. We investigate the effect of a variable brane tension on the cosmological evolution of the Universe, in the framework of a particular model in which the brane tension is an exponentially dependent function of the scale factor. The resulting cosmology shows the presence of an initial inflationary expansion, followed by a decelerating phase, and by a smooth transition towards a late accelerated de Sitter type expansion. The varying brane tension is also responsible for the generation of the matter in the Universe (reheating period). The physical constraints on the model parameters, resulted from the observational cosmological data, are also investigated.

\end{abstract}


\maketitle

\section{Introduction}

The idea of embedding our Universe in a higher dimensional space has
attracted a considerable interest recently, due to the proposal by Randall
and Sundrum that our four-dimensional (4D) spacetime is a three-brane,
embedded in a 5D spacetime (the bulk) ~\cite{Randall,Randall2}. This
proposal is based on early studies on superstring theory and M-theory, which
have suggested that our four dimensional world is embedded into a higher
dimensional spacetime. Particularly, the 10 dimensional $E_8 \otimes E_8$
heterotic superstring theory is a low-energy limit of the 11 dimensional
supergravity, under the compactification scheme $M^{10}\times S_1 / Z_2$
\cite{Witten,Witten2}. Thus, the 10 dimensional spacetime is compactified as
$M^4 \times CY^6 \times S_1 / Z_2$, implying that our Universe (a brane) is
embedded into a higher dimensional bulk. In this paradigm, the standard
model particles are open strings, confined on the braneworld, whilst the
gravitons and the closed strings can freely propagate into the bulk \cite
{Polchinski}.

The Randall-Sundrum Type II model has the virtue of providing a new type of
compactification of gravity \cite{Randall,Randall2}. Standard 4D gravity can
be recovered in the low-energy limit of the model, with a 3-brane of
positive tension embedded in 5D anti-de Sitter bulk. The covariant
formulation of the braneworld models has been formulated in \cite{Shiromizu}%
, leading to the modification of the standard Friedmann equations on the
brane. It turns out that the dynamics of the early Universe is altered by
the quadratic terms in the energy density and by the contribution of the
components of the bulk Weyl tensor, which both give a contribution in the
energy momentum tensor. This implies a modification of the basic equations
describing the cosmological and astrophysical dynamics, which has been
extensively considered recently \cite{all2}.

The recent observations of the CMB anisotropy by WMAP \cite{Spergel} have
provided convincing evidence for the inflationary paradigm \cite{Guth},
according to which in its very early stages the Universe experienced an
accelerated (de Sitter) expansionary phase (for recent reviews on inflation
see \cite{infl}).

At the end of inflation, the Universe is in a cold and low-entropy phase,
which is utterly different from the present hot high-entropy Universe.
Therefore the Universe should be reheated, or defrosted, to a high enough
temperature, in order to recover the standard Hot Big Bang \cite{reh}. The
reheating process may be envisioned as follows: the energy density in
zero-momentum mode of the scalar field decays into normal particles with
decay rate $\Gamma$. The decay products then scattered and thermalize to
form a plasma \cite{infl}.

Apart from the behavior of the inflaton field, the evolutions of dark energy
and dark matter in reheating stage were also considered. In \cite{Susperregi}%
, dark energy and dark matter were originated from a scalar field in
different stages of the inflation, according to a special form of potential.
Meanwhile, the conditions for unifying the description of inflation, dark
matter and dark energy were considered in \cite{Liddle3}. A specific model
was later proposed in \cite{Cardenas}, by using a modified quadratic scalar
potential. The candidates of dark matter in \cite{Liddle3} and \cite
{Cardenas} were oscillations of a scalar field. However, it may be possible
that dark matter existed on its own without originating from the scalar
field. This may pose less stringent constraint on the scalar field, so that
dark matter can be included in inflation paradigm in a easier way. On the
other hand, it was proposed that the decay products of scalar field acquired
thermal mass \cite{Kolb}.

The reheating in the braneworld models has also been considered recently. In
the context of the braneworld inflation driven by a bulk scalar field, the
energy dissipation from the bulk scalar field into the matter on the brane
was studied in \cite{HiTa03}. The obtained results supports the idea that
the brane inflation model, caused by a bulk scalar field, may be a viable
alternative scenario of the early Universe. The inflation and reheating in a
braneworld model derived from Type IIA string theory was studied in \cite
{BrDa03}. In this model the inflaton can decay into scalar and spinor
particles, thus reheating the Universe. A model in which high energy brane
corrections allow a single scalar field to describe inflation at early
epochs and quintessence at late times was discussed in \cite{SaDaSh03}. The
reheating mechanism in the model originates from Born-Infeld matter, whose
energy density mimics cosmological constant at very early times and
manifests itself as radiation subsequently. The particle production at the
collision of two domain walls in a 5-dimensional Minkowski spacetime was
studied in \cite{TaMa04}. This may provide the reheating mechanism of an
ekpyrotic (or cyclic) brane Universe, in which two BPS branes collide and
evolve into a hot big bang Universe. The reheating temperature $T_{\rm RH}$ in
models in which the Universe exits reheating at temperatures in the MeV
regime was studied in \cite{Hannestad}, and a minimum bound on $T_{\rm RH}$ was
obtained. The derived lower bound on the reheating temperature also leads to
very stringent bounds on the compactification scale in models with $n$ large
extra dimensions. The dark matter problem in the Randall-Sundrum type II
braneworld scenario was discussed in \cite{Pa05}, by assuming that the
lightest supersymmetric particle is the axino. The axinos can play the role
of cold dark matter in the Universe, due to the higher reheating
temperatures in the braneworld model, as compared to the conventional
four-dimensional cosmology. The impact of the non-conventional brane
cosmology on the relic abundance of non-relativistic stable particles in
high and low reheating scenarios was investigated in \cite{DaKh06}. In the
case of high reheating temperatures, the brane cosmology may enhance the
dark matter relic density by many order of magnitudes, and a stringent lower
bound on the five dimensional scale may be obtained. In the non-equilibrium
case, the resulting relic density is very small. The curvaton dynamics in
brane-world cosmologies was studied in \cite{PaZa06}.

Brane-worlds with non-constant tension, based on the analogy with fluid
membranes, which exhibit a temperature-dependence according to the empirical
law established by E\"otv\"os, were introduced in \cite{Ger1}. This new
degree of freedom allows for evolving gravitational and cosmological
constants, the latter being a natural candidate for dark energy. The
covariant dynamics on a brane with variable tension was studied in its full
generality, by considering asymmetrically embedded branes, and allowing for
non-standard model fields in the 5-dimensional space-time. This formalism
was applied for a perfect fluid on a Friedmann brane, which is embedded in a
5-dimensional charged Vaidya-Anti de Sitter space-time. For cosmological
branes a variable brane tension leads to several important consequences. A
variable brane tension may remove the initial singularity of the Universe,
since the brane Universe was created at a finite temperature $T_{c}$ and
scale factor $a_{\min}$ \cite{Ger2}. Both the brane tension and the
4-dimensional gravitational coupling 'constant' increase with the scale
factor from zero to asymptotic values. The 4-dimensional cosmological constant is
dynamical, evolving with $a$, starting with a huge negative value, passing
through zero, and finally reaching a small positive value. Such a
scale--factor dependent cosmological constant has the potential to generate
additional attraction at small $a$ (as dark matter does) and late-time
repulsion at large $a$ (dark energy). The evolution of the brane tension is
compensated by energy interchange between the brane and the fifth dimension,
such that the continuity equation holds for the cosmological fluid \cite{Ger2}. The
resulting cosmology closely mimics the standard model at late times, a
decelerated phase being followed by an accelerated expansion. The energy
absorption of the brane drives the 5D space-time towards maximal symmetry,
thus becoming Anti de Sitter. Other physical and cosmological implications of a varying brane tension have been considered in \cite{yun}.

%Thus, our model is different from the model presented in
%\cite{Umezu}, where scalar fields are absent, and the injection of
%the dark matter from the bulk to the brane is used to explain the
%cosmic acceleration. The physical constraints used in \cite{Umezu}
%also differ from those used in this paper as well. For instance,
%instead  of using the CMB temperature fluctuations and Type Ia
%supernovae at high redshift.

%The present paper is organized as follows.  The basic equations of
%our model are formulated in Section II. A dimensionless form of
%the field equations is presented in Section III, where the
%physically acceptable range of the values of the physical
%parameters is also discussed. The extra-dimensional effects
%dominated phase of the reheating, during which the quadratic terms
%in the energy density dominate the dynamics and evolution of the
%Universe, is discussed in Section IV. The reheating process is
%considered in its full generality in Section V, and the role of
%the linear terms in the energy density is investigated. Finally,
%in Sec. VI we summarize and conclude our results.

It is the purpose of the present paper to further investigate the cosmological implications of a varying brane tension. As a first step in our study, we consider a thermodynamic interpretation of the varying brane tension models,  by showing that the field equations with variable $\lambda $ can be interpreted as describing matter creation in a cosmological framework. The particle creation rate is determined by the variation rate of the brane tension, as well as by the brane-bulk energy-matter transfer rate. In particular, by adopting a theoretical model in which the brane tension is a simple function of the scale factor of the Universe, we consider the possibility that the early inflationary era in the evolution of the brane Universe was driven by a varying brane tension. A varying brane tension may also be responsible for the generation of the matter after reheating, as well as for the late time acceleration of the Universe.

The present paper is organized as follows. In Section~\ref{geo} we present the field equations of the brane world models with varying brane tension and we write down the basic equations describing the cosmological dynamics of a flat Friedmann-Robertson-Walker Universe. The thermodynamic interpretation of the brane-world models with varying brane tension and brane-bulk matter-energy exchange is considered in Section~\ref{therm}. A power-law inflationary brane-world model with varying brane tension and non-zero bulk pressure is obtained in Section~\ref{pow}. The analytical behavior of the cosmological model with varying brane tension is considered in Section~\ref{scale}, by using the small and large time approximations for the brane tension. The numerical analysis of the model is performed in Section~\ref{numerical}. We discuss and conclude our results in Section~\ref{conclusion}.

\section{Geometry and field equations in the variable brane tension models}\label{geo}

In the present Section we present the field equations for brane world models with varying brane tension, and the corresponding cosmological field equations for a flat Robertson-Walker space-time.

\subsection{Gravitational field equations}

We start by considering a five dimensional ($5D$) spacetime (the bulk), with a large negative 5D cosmological constant ${}^{(5)}\Lambda$ and a single four-dimensional ($4D$) brane, on which usual (baryonic) matter and
physical fields are confined. The $4D$ braneworld $({}^{(4)}M,{}^{(4)}g_{\mu \nu })$
is located at a hypersurface $\left(B\left( X^{A}\right) =0\right)$ in the $%
5D$ bulk spacetime $({}^{(5)}M,{}^{(5)}g_{AB})$ with mirror symmetry, and with coordinates
$X^{A},A=0,1,...,4$. The induced $4D$ coordinates on the brane are $%
x^{\mu },\mu =0,1,2,3$. We choose normal Gaussian coordinates, and therefore the $5D$ metric is related to the $4D$ metric by the relation ${}^{(5)}g_{MN}={}^{(4)}g_{MN}+n_{M}n_{N}$, where $n^{M}$ is the normal vector.

The induced $4D$ metric is $g_{IJ}={}^{(5)}g_{IJ}-n_{I}n_{J}$, where $n_{I}$
is the space-like unit vector field normal to the brane hypersurface $%
{}^{(4)}M$. The basic equations on the brane are obtained by projections
onto the brane world with Gauss equation, Codazzi equation and Israel
junction condition, the projected Einstein equation are given by
\begin{equation}
G_{\mu \nu }=-\Lambda g_{\mu \nu }+k^{2}T_{\mu \nu }+\bar{k}^{4}S_{\mu \nu }-%
\bar{\epsilon}_{\mu \nu }+\bar{L}_{\mu \nu }^{TF}+\bar{P}_{\mu \nu }+F_{\mu \nu },
\end{equation}
where
\begin{equation}
S_{\mu \nu }=\frac{1}{2}TT_{\mu \nu }-\frac{1}{4}T_{\mu \alpha }T_{\nu
}^{\alpha }+\frac{3T_{\alpha \beta }T^{\alpha \beta }-T^{2}}{24}g_{\mu \nu },
\end{equation}
\begin{equation}\label{eps}
\varepsilon _{\mu \nu }=C_{ABCD}n^{C}n^{D}g_{\mu }^{A}g_{\nu }^{B},
\end{equation}
and
\begin{equation}
F_{\mu \nu }=^{(5)}T_{AB}g_{\mu }^{A}g_{\nu }^{B}+\left(
^{(5)}T_{AB}n^{A}n^{B}-\frac{1}{4}^{(5)}T\right) g_{\mu \nu },
\end{equation}
respectively.

Apart from the terms quadratic in the brane energy-momentum tensor, in the
field equations on the brane there are two supplementary terms,
corresponding to the projection of the $5D$ Weyl tensor $\varepsilon _{\mu
\nu }$ and of the projected tensor $F_{\mu \nu }$, which contains the bulk
matter contribution. Both terms induce bulk effects on the brane.

Also, the possible asymmetric embedding is characterized by the tensor
\begin{equation}
\bar{L}_{\mu \nu }=\bar{K}_{\mu \nu }\bar{K}-\bar{K}_{\mu \sigma }\bar{K}%
_{\nu }^{\sigma }-\frac{g_{\mu \nu }}{2}\left( \bar{K}^{2}-\bar{K}_{\alpha
\beta }\bar{K}^{\alpha \beta }\right) ,
\end{equation}
with trace
$\bar{L}=\bar{K}_{\alpha \beta }\bar{K}^{\alpha \beta }-\bar{K}^{2}$,
and trace-free part
$\bar{L}_{\mu \nu }^{TF}=\bar{K}_{\mu \nu }\bar{K}-\bar{K}_{\mu \sigma }\bar{K%
}_{\nu }^{\sigma }+\bar{L}g_{\mu \nu }/4$, respectively.

For a $Z_2$ symmetric embedding $\bar{K}_{\mu \nu }=0$, and thus $\bar{L}%
_{\mu \nu }=0$. $\bar{P}_{\mu \nu }$ is given by the pull-back to the brane
of the energy-momentum tensor characterizing possible non-standard model
fields (e. g. scalar, dilaton, moduli, radiation of quantum origin) living
in 5D,
\begin{equation}
\bar{P}_{\mu \nu }=\frac{2\tilde{k}^{2}}{3}\overline{\left( g_{\mu }^{\alpha
}g_{\nu }^{\beta }{}^{(5)}T_{\alpha \beta }\right) ^{TF}},
\end{equation}
which is traceless by definition. Another projection of the 5D sources
appears in the brane cosmological constant $\Lambda $. which is defined as
\begin{equation}
\Lambda =\Lambda _{0}-\frac{\bar{L}}{4}-\frac{2\tilde{k}^{2}}{3}\overline{%
\left( n^{\alpha }n^{\beta }{}^{(5)}T_{\alpha \beta }\right)},
\end{equation}
where $2\Lambda _{0}=k_5^{2}\lambda +k_5^{2}\Lambda_5$.

In the case of a variable brane tension, the projected gravitational field
equations on the brane have a similar form to the general case,
\begin{equation}
G_{\mu \nu }=-\Lambda g_{\mu \nu }+k^{2}T_{\mu \nu }+\bar{k}^{4}S_{\mu \nu }-%
\bar{\epsilon}_{\mu \nu }+\bar{L}_{\mu \nu }^{TF}+\bar{P}_{\mu \nu }+F_{\mu \nu }.
\end{equation}
However, the evolution of the brane tension appears in the Codazzi equation, and in
the differential Bianchi identity. The Codazzi equation is
\begin{equation}
\nabla_{\mu}\bar{K^{\mu}_{\nu}}-\nabla_{\nu}\bar{K}=k_5^2\overline{%
(g^{\rho}_{\nu}n^{\sigma}{}^{(5)}T_{\rho\sigma})},
\end{equation}
and it gives the conservation equation of the matter on the brane as
\begin{equation}  \label{codazzi}
\nabla_{\mu}T^{\mu}_{\nu}=\nabla_{\nu}\lambda-\Delta(g^{\rho}_{\nu}n^{%
\sigma}{}^{(5)}T_{\rho\sigma}).
\end{equation}
The differential Bianchi identity, written as
$\nabla^{\mu}R_{\rho\mu}=\frac{1}{2}\nabla_{\rho}R$, gives
\begin{eqnarray}  \label{2bi}
\nabla^{\mu}(\bar{\epsilon}_{\mu\nu}-\overline{L}^{TF}_{\mu\nu}-\bar{%
\mathcal{P}}_{\mu\nu})& =\frac{\nabla_{\nu}\bar{L}}{4}+\frac{k_5^2}{2}%
\nabla_{\nu}\overline{(n^{\rho}n^{\sigma}{}^{(5)}T_{\rho\sigma})}-\frac{%
k_5^4\lambda}{6}\Delta(g^{\sigma}_{\nu}n^{\rho}T_{\sigma\rho}) \nonumber\\
& +\frac{k_5^4}{4}(T^{\mu}_{\nu}-\frac{T}{3}g^{\mu}_{\nu})\Delta(g^{%
\sigma}_{\mu}n^{\rho}{}^{(5)}T_{\sigma\rho})+\frac{k_5^4}{4}%
[2T^{\mu\sigma}\nabla_{[\nu}T_{\mu]\sigma} \nonumber\\
& +\frac{1}{3}(T_{\mu\nu}\nabla^{\mu}T-T\nabla_{\nu}T)]-\frac{k_5^4}{12}%
(T^{\mu}_{\nu}-Tg^{\mu}_{\nu})\nabla_{\mu}\lambda).
\end{eqnarray}

From  Eq.~(\ref{eps}), one can introduce an effective non-local energy density $U$, which can be obtained by
assuming that $\varepsilon_{\mu\nu}$  in the projected Einstein equation
behaves as an effective radiation fluid,
\begin{equation}
-\varepsilon _{\mu \nu }=%
\frac{k_{5}^{4}}{6}\lambda U(u_{\mu }u_{\nu }+\frac{a^{2}}{3}h_{\mu \nu }),
\end{equation}
where $u_{\mu }$ is the matter four-velocity, and $h_{\mu \nu}=g_{\mu \nu}+u_{\mu \nu }$, respectively.

\subsection{Cosmological models with dynamic brane tension}\label{cosmo}

We assume that the metric on the brane is given by the flat Robertson-Walker-Friedmann metric, with
\begin{equation}
{}^{(4)}g_{\mu\nu}dx^{\mu}dx^{\nu}=-dt^2+a^2(t)(dx^2+dy^2+dz^2),
\end{equation}
where $a$ is the scale factor. The matter on the brane is assumed to consist of a perfect fluid, with energy density $\rho$, and pressure $p$, respectively.
The gravitational field equations, governing the evolution of the brane Universe with variable brane tension, in the presence of brane-bulk energy transfer, and with a non-zero bulk pressure, are then given by \cite{Ger1, Ger2}

\begin{eqnarray}\label{feq}
\left( \frac{\dot{a}}{a}\right) ^{2} &=&\frac{{}^{(4)}\Lambda }{3}+\frac{%
k_{5}^{4}\lambda }{18}\left[ \rho +\frac{\rho ^{2}}{2\lambda }+U\right] ,\label{1} \\
\frac{\ddot{a}}{a} &=&\frac{{}^{(4)}\Lambda }{3}-\frac{k_{5}^{4}\lambda }{36}%
\left[ \rho \left( 1+\frac{2\rho }{\lambda }\right) +3p\left( 1+\frac{\rho }{%
\lambda }\right) +2U\right] , \label{2}\\
\dot{\rho}+3H\left( \rho +p\right) &=&-\dot{\lambda}-2P_{5}, \label{3}\\
\frac{k_{5}^{4}\lambda }{6}\left( \dot{U}+4U\frac{\dot{a}}{a}+U\frac{\dot{%
\lambda}}{\lambda }\right) &=&\frac{k_{5}^{2}}{2}\dot{\bar{P}}_{B}+\frac{%
k_{5}^{4}\lambda }{3}\left( 1+\frac{\rho }{\lambda }\right) P_{5}, \label{4}\\
{}^{(4)}\Lambda &=&\frac{k_{5}^{2}}{2}\Lambda _{5}+\frac{k_{5}^{4}}{12}%
\lambda ^{2}-\frac{k_{5}^{2}}{2}\bar{P}_{B},\label{5}
\end{eqnarray}
where $P_5$ describes the bulk-brane matter-energy transfer, while $P_B$ is the bulk pressure.
%\begin{eqnarray}\label{feq}
%\left( \frac{\dot{a}}{a}\right) ^{2} &=&\frac{{}^{(4)}\Lambda }{3}+\frac{%
%k_{5}^{4}\lambda }{18}\left[ \rho +\frac{\rho ^{2}}{2\lambda }+U\right] , \label{1}\\
%\frac{\ddot{a}}{a} &=&\frac{{}^{(4)}\Lambda }{3}-\frac{k_{5}^{4}\lambda }{36}%
%\left[ \rho \left( 1+\frac{2\rho }{\lambda }\right) +3p\left( 1+\frac{\rho }{%
%\lambda }\right) +2U\right] , \label{2}\\
%\dot{\rho}+3H\left( \rho +p\right) &=&-\dot{\lambda}, \label{3}\\
%\frac{k_{5}^{4}\lambda }{6}\left( \dot{U}+4U\frac{\dot{a}}{a}+U\frac{\dot{%
%\lambda}}{\lambda }\right) &=&0, \label{4}\\
%{}^{(4)}\Lambda &=&\frac{k_{5}^{2}}{2}{}^{(5)}\Lambda+\frac{k_{5}^{4}}{12}%
%\lambda ^{2}.\label{5}
%\end{eqnarray}

An important observational parameter, which is an indicator of the
rate of expansion of the Universe, is the deceleration parameter $q$, defined as
\begin{equation}
q=\frac{d}{dt}\left( \frac{1}{H}\right) -1=-\frac{a\ddot{a}}{\dot{a}^{2}}=-\frac{\ddot{a}/a}{\left(\dot{a}/a\right)^2}.
\label{q0}
\end{equation}
If $q<0$, the expansion of the Universe is accelerating, while $q>0$ indicates a decelerating phase.

\section{Thermodynamic interpretation of the varying tension in brane-world models}\label{therm}

For the sake of generality we also assume that there is an effective energy-matter transfer between the brane and
the bulk, and the brane-bulk matter-energy exchange can be described as
\begin{equation}\label{bbtrans}
P_{5}=-\frac{\alpha _{bb}}{2}\rho _{cr}\left( \frac{a_{0}}{a}\right) ^{3w}H,
\end{equation}%
where $\alpha _{bb}$ is a constant, $\rho _{cr}$ is the present day critical
density of the Universe, and $a_{0}$ is the present day value of the scale
factor.

In the presence of a varying brane tension and of the bulk-brane matter and
energy exchange, the energy conservation equation on the brane can be
written as
\begin{equation}
\dot{\rho}+3(\rho +p)H=-\rho \left( \frac{\dot{\lambda}}{\rho }-\alpha
_{bb}H\right) ,  \label{partcreat}
\end{equation}%
where we have used Eq.~(\ref{bbtrans}) for the description of brane bulk
energy transfer, by taking into account that $\rho =\rho _{cr}\left(
a_{0}/a\right) ^{3w}$. We suppose that the matter content of the early
Universe is formed from $m$ non-interacting comoving relativistic fluids
with energy densities and thermodynamic pressures $\rho _{i}(t)$ and $%
p_{i}(t)$, $i=1,2,...,m$, respectively, with each fluid formed from
particles having a particle number density $n_{i}(t)$, $i=1,2,...,m$, and
obeying equations of state of the form $\rho _{i}(t)=k_{i}n_{i}^{\gamma _{i}}
$, $p_{i}(t)=\left( \gamma _{i}-1\right) \rho _{i}$, $i=1,2,...,m$, where $%
k_{i}=\rho _{0i}/n_{0i}^{\gamma _{i}}\geq 0$, $i=1,2,...,m$, are constants,
and $1\leq \gamma _{i}\leq 2$, $i=1,2,...,m$. For example, we can consider
that the particle content of the early Universe is determined by pure
radiation (i. e., different types of massless particles, or massive matter
(baryonic and dark) in equilibrium with electromagnetic radiation and
decoupled massive particles. The total energy density and pressure of the
cosmological fluid results from summing the contribution of the $l$ simple
fluid components, and are given by $\rho \left( t\right) =\sum_{i=1}^{m}\rho
_{i}(t)$ and $p\left( t\right) =\sum_{i=1}^{m}p_{i}(t)$, respectively. For a
multicomponent comoving cosmological fluid and in the presence of variable
brane tension and bulk-brane energy exchange, Eq.~(\ref{partcreat}) becomes
\begin{equation}
\sum_{i=1}^{l}\left[ \dot{\rho}_{i}+3(\rho _{i}+p_{i})H\right]
=-\sum_{i=1}^{m}\rho _{i}(t)\left[ \frac{\dot{\lambda}}{\sum_{i=1}^{m}\rho
_{i}(t)}-\alpha _{bb}H\right] .  \label{partcreat1}
\end{equation}

Eq.~(\ref{partcreat1}) can be recast into the form of $m$ particle balance
equations,
\begin{equation}
\dot{n}_{i}(t)+3n_{i}(t)H=\Gamma _{i}(t)n_{i}(t),i=1,2,...,m,
\label{partcreat2}
\end{equation}%
where $\Gamma _{i}(t)$, $i=1,2,...,m$, are the particle production rates,
given by
\begin{equation}
\Gamma _{i}(t)=-\frac{1}{\gamma _{i}}\left[ \frac{\dot{\lambda}}{m\rho
_{i}(t)}-\alpha _{bb}H\right] ,i=1,2,...,m.  \label{Gamma}
\end{equation}

In order for Eq.~(\ref{partcreat2}) to describe particle production the
condition  $\Gamma _{i}(t)\geq 0$, $i=1,2,...,m$, is required to be
satisfied, leading to the following restriction imposed to the time
variation rate of the brane tension
\begin{equation}
\dot{\lambda}\leq \alpha _{bb}l\rho _{i}(t)H,i=1,2,...,m.
\end{equation}

Note that if $\Gamma _{i}(t)=0$, $i=1,2,...,m$, we obtain the usual particle
conservation law of the standard cosmology. Of course, the casting of Eq.~(%
\ref{partcreat1}) is not unique. In Eqs.~(\ref{partcreat2}) and (\ref{Gamma}%
), we consider the simultaneous creation of a multicomponent comoving
cosmological fluid, but other possibilities can be formulated in the same
way (for example, creation of a single component in a mixture of fluids).

The entropy $S_{i}$,  generated during particle creation at
temperatures $T_{i}$, $i=1,2,...,m$, can be obtained from Eq.~(\ref%
{partcreat2}), and for each species of particles has the expression
\begin{equation}
T_{i}\frac{dS_{i}}{dt}=-\frac{1}{\gamma _{i}}\left[ \frac{\dot{\lambda}}{%
m\rho _{i}(t)}-\alpha _{bb}H\right] \rho _{i}\left( t\right) V,i=1,2,...,m,
\end{equation}%
where $V$ is the volume of the Universe, or, equivalently,
\begin{equation}
\frac{dS_{i}}{dt}=\frac{\gamma _{i}\rho _{i}(t)V}{T_{i}}\Gamma
_{i}(t),i=1,2,...,m.
\end{equation}

In a cosmological fluid where the density and pressure are functions of the
temperature only, $\rho =\rho \left( T\right) $, $p=p\left( T\right) $, the
entropy of the fluid is given by $S=\left( \rho +p\right) V/T=\gamma \rho
\left( t\right) V/T$. Therefore we can express the total entropy $S(t)$ of
the multicomponent cosmological fluid filled brane Universe as a function of
the particle production rate only,
\begin{equation}
S(t)=\sum_{i=1}^{m}S_{0i}\exp \left[ \int_{t_{0}}^{t}\Gamma _{i}\left(
t^{\prime }\right) dt^{\prime }\right] ,
\end{equation}%
where $S_{0i}\geq 0$, $i=1,2,...,m$, are constants of integration. In the
case of a general perfect comoving multicomponent cosmological fluid with
two essential thermodynamical variables, the particle number densities $n_{i}
$, $i=1,2,...,m$, and the temperatures $T_{i}$, $i=1,2,...,m$, it is
conventional to express $\rho _{i}$ and $p_{i}$ in terms of $n_{i}$ and $%
T_{i}$ by means of the equilibrium equations of state $\rho _{i}=\rho
_{i}\left( n_{i},T_{i}\right) $, $p_{i}=p_{i}\left( n_{i},T_{i}\right) $, $%
i=1,2,...,m$. By using the general thermodynamic relation
\begin{equation}
\frac{\partial \rho _{i}}{\partial n_{i}}=\frac{\rho _{i}+p_{i}}{n_{i}}-%
\frac{T_{i}}{n_{i}}\frac{\partial p_{i}}{\partial T_{i}},i=1,2,...,m,
\end{equation}%
in the case of a general
comoving multicomponent cosmological fluid Eq.~(\ref{partcreat}) can also be rewritten  in the form of $m$ particle
balance equations,
\begin{equation}
\dot{n}_{i}(t)+3n_{i}(t)H=\Gamma _{i}(t)n_{i}(t),i=1,2,...,m,
\label{partcreat3}
\end{equation}%
with the particle production rates $\Gamma _{i}(t)$ given by some
complicated functions of the thermodynamical parameters, brane tension and
brane-bulk energy exchange rate,
\begin{equation}
\Gamma _{i}(t)=-\frac{\rho _{i}}{\rho _{i}+p_{i}}\left[ \frac{\dot{\lambda}}{%
l\rho _{i}(t)}-\alpha _{bb}H+T_{i}\frac{\partial \ln \rho _{i}}{\partial
T_{i}}\left( \frac{\dot{T}_{i}}{T_{i}}-C_{i}^{2}\frac{\dot{n}_{i}}{n_{i}}%
\right) \right] ,i=1,2,...,m,
\end{equation}%
where $C_{i}^{2}=\left( \partial p_{i}/\partial T_{i}\right) /\left(
\partial \rho _{i}/\partial T_{i}\right) $. The requirement that the
particle balance equation Eq.~(\ref{partcreat3}) describes particle
production,  $\Gamma _{i}(t)\geq 0$, $i=1,2,...,m$, imposes in this case the
following constraint on the time variation of the brane tension,
\begin{equation}
\dot{\lambda}<\alpha _{bb}l\rho _{i}(t)H-l\rho _{i}(t)T_{i}\frac{\partial
\ln \rho _{i}}{\partial T_{i}}\left( \frac{\dot{T}_{i}}{T_{i}}-C_{i}^{2}%
\frac{\dot{n}_{i}}{n_{i}}\right), i=1,2,...,m .
\end{equation}

In the general case the entropy generated during the reheating period due to
the variation of the brane tension and the bulk-brane energy exchange can
be obtained for each component of the cosmological fluid from the equations
\begin{equation}
\frac{dS_{i}}{dt}=\frac{\left( \rho _{i}+p_{i}\right) V}{T_{i}}\left[ \frac{%
\dot{\lambda}}{l\rho _{i}(t)}-\alpha _{bb}H+T_{i}\frac{\partial \ln \rho _{i}%
}{\partial T_{i}}\left( \frac{\dot{T}_{i}}{T_{i}}-C_{i}^{2}\frac{\dot{n}_{i}%
}{n_{i}}\right) \right] ,i=1,2,...,m,
\end{equation}%
while the total entropy of the Universe is given by $S(t)=%
\sum_{i=1}^{m}S_{i}(t)$.

The entropy flux vector of the $k$th component of the cosmological fluid is
given by
\begin{equation}
S^{(k)\alpha }=n_{k}\sigma _{k}u^{\alpha },k=1,2,..,m,
\end{equation}%
where $\sigma _{k}$, $k=1,2,...,m$, is the specific entropy (per particle)
of the corresponding cosmological fluid component and $u^{\alpha }$ is the
four-velocity of the fluid.  By using the Gibbs equation $nTd\sigma =d\rho -%
\left[ \left( \rho +p\right) /n\right] dn$ for each component of the fluid,
and assuming that the entropy density $\sigma $ does not depend on the brane
tension, we obtain
\begin{equation}
S_{;\alpha }^{(k)\alpha }=-\frac{1}{T_{k}}\left( \dot{\lambda}-\alpha
_{bb}H\rho \right) -\frac{\mu _{k}\Gamma _{k}n_{k}}{T_{k}},k=1,2,..,m,
\end{equation}%
where $\mu _{k}$ is the chemical potential defined by $\mu _{k}=\left[
\left( \rho _{k}+p_{k}\right) /n_{k}\right] -T_{k}\sigma _{k}$. The chemical
potential is zero for radiation. For each component of the cosmological
fluid the second law of thermodynamics requires that the condition
\begin{equation}
S_{;\alpha }^{(k)\alpha }\geq 0,k=1,2,..,m,
\end{equation}%
has to be satisfied.

\section{Power law inflation in brane world models with varying brane tension and bulk pressure}\label{pow}

For a vacuum Universe
with $\rho =p=0$, in the presence of a non-zero bulk pressure and matter-
energy exchange between the brane and the bulk, the field equations Eqs. (%
\ref{feq}) take the form
\begin{equation}\label{infl1}
3\left( \frac{1}{a}\frac{da}{d\tau }\right) ^{2}=\frac{l^{2}}{2}-p_{B}+lu,
\end{equation}
\begin{equation}\label{infl2}
3\frac{1}{a}\frac{d^{2}a}{d\tau ^{2}}=\frac{l^{2}}{2}-p_{B}-lu,
\end{equation}
\begin{equation}\label{infl3}
\frac{dl}{d\tau }=-2\sqrt{\frac{2}{3}}p_{5},
\end{equation}%
and
\begin{equation}\label{infl4}
\frac{d}{d\tau }\left( lua^{4}\right) =-a^{4}\frac{d}{d\tau }\left( \frac{%
l^{2}}{2}-p_{B}\right) ,
\end{equation}%
where we have introduced a set of dimensionless variables $\left( \tau
,l,p_{B},u,p_{5}\right) $ defined as
\begin{equation}
\tau =\sqrt{\frac{3}{2}}t,\lambda =\frac{3}{k_{5}^{2}}l,P_{B}=\frac{3}{%
k_{5}^{2}}p_{B},U=\frac{3}{k_{5}^{2}}u,P_{5}=\frac{3}{k_{5}^{2}}p_{5}.
\end{equation}
Moreover, we consider that the five-dimensional cosmological constant $\Lambda _5=0$. We assume that the inflationary evolution is of the power law type, and
therefore $a=\tau ^{\alpha }$, where $\alpha $ is a constant. Then Eqs. (\ref%
{infl1}) and (\ref{infl2}) give

\begin{equation}
2\left(\frac{l^{2}}{2}-p_{B}\right)=\frac{3\alpha \left( 2\alpha -1\right) }{\tau ^{2}},
\end{equation}%
and
\begin{equation}
2lu=\frac{3\alpha }{\tau ^{2}}.
\end{equation}

Eq.~(\ref{infl4}) is identically satisfied. In order to completely solve the
problem, we need to specify the form of the energy matter-transfer from the
bulk to the brane. By assuming a functional form given by $p_{5}=p_{05}\tau
^{-\beta }$, where $\beta >0$ and $p_{05}>0$ are constants, we obtain
immediately
\begin{eqnarray}
&&l\left( \tau \right)=\sqrt{\frac{8}{3}}\frac{1}{\beta -1}\tau ^{-\beta
+1},p_{B}\left( \tau \right) =\frac{4}{3}\frac{1}{\left( \beta -1\right) ^{2}%
}\tau ^{-2(\beta -1)}-\frac{3\alpha \left( 2\alpha -1\right) }{2\tau ^{2}},\nonumber\\
&&u\left( \tau \right) =\sqrt{\frac{27}{32}}\alpha \left( \beta -1\right)
\tau ^{\beta -3}.
\end{eqnarray}
The Hubble parameter of the Universe during the inflationary phase is given by $H=\alpha /t$. The deceleration parameter is obtained as $q=d\left(1/H\right)/dt-1=\left(1-\alpha\right)/\alpha $. Therefore, if $\alpha >1$, $q<0$, and the brane world Universe experiences an inflationary expansion.

\section{Scale factor dependent brane tension models}\label{scale}

In the following we assume that there is no matter-energy exchange between the bulk and the brane, $P_5=0$, and that the bulk pressure is also zero, $P_B=0$. For the matter on the brane we adopt as equation of state a linear barotropic relation between density and pressure, given by
\begin{equation}
p=\left( w-1\right) \rho ,
\end{equation}
where $w={\rm constant}$ and $w\in \lbrack 1,2]$. Therefore Eq.~(\ref{3}) gives
\begin{equation}
\dot{\rho}+3Hw\rho =-\dot{\lambda},
\label{3w}
\end{equation}
while Eq.~(\ref{4}) gives immediately
\begin{equation}
\lambda U=\frac{U_{0}}{a^{4}},
\end{equation}
where $U_{0}$ is an arbitrary constant of integration. In the following, in order to simplify the analysis, we assume that $U_0=0$.

In order to explain the main observational features of modern cosmology (inflation, reheating, deceleration period and late time acceleration, respectively), we assume that the brane tension varies as a function of the scale factor $a$ according to the equation
\begin{equation}
\lambda^2=\lambda_0^2e^{-2\beta a^2}-\frac{6{}^{(5)}\Lambda}{k_5^2}+\lambda_1^2,
\label{vbrane}
\end{equation}
where $\beta $, $\lambda _0$ and $\lambda _1$ are constants.
%It will be shown in the later section that the value of $\lambda_0$ is fixed by the time scale of inflation and the value of $\beta$ is fixed by the e-folding of inflation adopted.
%This is an phenomenological choice of brane tension so that it is steady at the very beginning and quickly decay to a constant. The constant is fine tuned to cancel the 5D cosmological term and remain a residue term $\lambda_1^2$.

Suppose $t_{\rm in}, a_{\rm in}$ and $\rho_{\rm in}$ are the values of the time, of the scale factor, and of the energy density before inflation. Generally, in the present paper we use the subscript $``{\rm in}"$ to denote the values of the cosmological parameters before the inflation, and the subscript $``{\rm en}"$ to denote values after inflation. Thus, for example,  $N=\ln\left(a_{\rm en}/a_{\rm in}\right)$ is the e-folding number. The basic physical parameters of our model are $t_{\rm in}$, $a_{\rm in}$, $\rho_{\rm in}$, $N, k_5$, ${}^{(5)}\Lambda$, $\lambda_0$, $\lambda_1 $, and $\beta$, respectively. The coupling constant $k_5$ and the five-dimensional cosmological constant ${}^{(5)}\Lambda$ are constrained by the present value of the gravitational constant,
\begin{equation}
\frac{k_5^4}{6}\sqrt{-\frac{6{}^{(5)}\Lambda}{k_5^2}+\lambda_1^2}\approx k_5^3\sqrt{-\frac{{}^{(5)}\Lambda}{6}}=8\pi G\approx 1.68\times10^{-55}{\rm eV}^{-2},
\label{gconstant}
\end{equation}
and by the constraints on the 5D cosmological constant \cite{Maartens}
\begin{equation}
\frac{k_5^2}{2}{}^{(5)}\Lambda\approx -\frac{6}{(0.1{\rm mm})^2}\approx -2.3\times 10^{-5}{\rm eV}^{2},
\end{equation}
where we have used the natural system of units with  $\hbar=c=1$. >From these two conditions, we obtain $k_5^4\approx 3.6\times 10^{-105}\;{\rm eV}^{-6}$ and ${}^{(5)}\Lambda\approx -7.7\times 10^{46}\;{\rm eV}^5$. Besides, the value of $\lambda_1$ can be obtained from the value of the present day dark energy $\rho_{\rm dark}\approx 10^{-12}{\rm eV}^4$ \cite{Carroll},
\begin{equation}
\frac{k_{5}^{4}}{12}\lambda_1 ^{2}=8\pi G \rho_{\rm dark}\approx 1.6\times10^{-67}{\rm eV}^2,
\label{matchdark}
\end{equation}
which gives $\lambda_1\approx 1.6\times10^{19}\;{\rm eV}^4$. We can have a backward checking on Eq.~(\ref{gconstant}), from which it follows that the condition $\lambda_1^2\ll -6{}^{(5)}\Lambda/k_5^2$ is indeed satisfied.
%relative error of $G$ ($6.7\times 10^{-57}{\rm eV}^{-2}$) is of $1.28\times10^{-4}$.
We also choose $\lambda_0$ to be of the same order of magnitude as the vacuum energy $\rho_{\rm vac}\backsim 10^{100}\;{\rm eV}^4$ at GUT scale \cite{Carroll}, \cite{Ishak},
\begin{equation}
\frac{k_{5}^{4}}{12}\lambda_0 ^{2}=8\pi G \rho_{\rm vac}\backsim 10^{45}{\rm eV}^2,
\label{matchvac}
\end{equation}
which gives $\lambda_0\backsim 1.3\times10^{75}$ eV. The scale difference between $\lambda_0$ and $\sqrt{-6{}^{(5)}\Lambda/k_5^2+\lambda_1^2}$ is $\lambda_0/\sqrt{-6{}^{(5)}\Lambda/k_5^2}\backsim 10^{25}$. The differences in the scales of $\lambda_0$ and $\lambda_1$ are of the order of $\backsim 56$. For the remaining model parameters $t_{\rm in}, a_{\rm in}, \rho_{\rm in}, N, \beta$, we constraint them in the next section.
%If we express the theoretical and the observed values of the vacuum energy in terms of the fourth power of the mass scale, $M_{\rm vac}^{\rm (theory)}\backsim 10^{18}\;{\rm eV}$ and $M_{\rm vac}^{\rm (obs)}\approx 6\times 10^4{\rm eV}$, respectively, they differ by order of $14$.

When $a$ is very small, the brane tension $\lambda\approx\lambda_0$ dominates the early Universe at the time of inflation. Due to the exponential expansion of the Universe, the brane tension quickly decays to a constant just after the inflation. The decay of the brane tension will generate the matter content of the Universe, according to Eq.~(\ref{3}). This happens also during the accelerated expansion period of the Universe. Matter is created during all periods of the expansion of the Universe, but the most important epoch for matter creation is near the end of inflation. In the evolution of the Universe there is one moment when $\ddot{a}=0$, which corresponds to the moment when the Universe switches from the accelerating expansion to a decelerating phase. After the matter (which is mainly in the form of radiation) energy density reaches its maximum, the Universe enters into a radiation dominated phase, and the quadratic term in Eq.~(\ref{1}) will become dominant first. The matter energy density continue to decrease due to the expansion. When the linear term in matter equals the quadratic term, the Universe switches back to the $\rm\Lambda CDM$ model. Therefore, the Universe enters in the matter dominated epoch at about $4.7\times10^4 \rm yr$ \cite{Ryden}. Then the matter term equals the residue term in Eq.~(8) at about $10\rm Gyr$ \cite{Ryden}. This is the second moment in the evolution of the Universe when $\ddot{a}=0$. After this moment, the Universe enters in an accelerating expansionary phase again, and its dynamics is controlled by the term $\lambda_1$.

\section{Qualitative analysis of the model}

In the present Section we consider the approximate behavior of the cosmological model with varying brane tension in the different cosmological epochs.

\subsection{Early inflationary phase: $2\beta a^2\ll 1$}

When the scale factor $a$ is very small, the exponential factor $e^{-2\beta a^2}$ in Eq.~(\ref{vbrane}) can be approximated by $1$. Therefore, the brane tension is given by $\lambda^{2}\approx \lambda_{0}^{2}$, and physically it corresponds to the vacuum energy necessary to give an exponential inflation.
%According to current quantum field theory the value of the vacuum energy is of the order of $\rho_{\rm vac}\approx 10^{72}\;{\rm GeV^4}$ \cite{infl}. Since  $\lambda_0$ should have the same value as this energy scale, we find $\lambda_0\approx $.
Since $k^2_5\lambda_{0}^2/6\gg{}^{(5)}\Lambda$, from Eq. (\ref{1}) the scale factor evolves in time as an exponential function of time, given by
\begin{equation}
a=a_{\rm in}e^{(k_5^2\lambda_0/6)\,t},
\label{approxa}
\end{equation}
where $a_{{\rm in}}$ is the value of the scale factor prior to inflation. The e-folding number is given by $N=\ln(a/a_{{\rm in}})$, which should be roughly of the order of $N\gtrsim 60$ in order to solve the flatness, Horizon problem, etc. In the present paper we adopt for $N$ the value $N=70$. Since $2\beta a_{\rm en}^2\backsim 1$ at the end of inflation, we can roughly estimate the value of $a_{\rm en}$ to be
\begin{equation}
a_{\rm en}\backsim \frac{1}{\sqrt{2\beta}}.
\label{aen}
\end{equation}
From the value of $N$ we adopted, we obtain the value of $a_{\rm in}$ as
\begin{equation}
a_{\rm in}=a_{\rm en}e^{-N}.
\end{equation}
According to Eq.~(\ref{approxa}), the end time of the inflation $t_{\rm en}$ can be estimated as
\begin{equation}
k_5^2\lambda_0(t_{\rm en}-t_{\rm in})/6\approx k_5^2\lambda_0 t_{\rm en}/6\approx N,
\end{equation}
which implies that $t_{\rm en}\approx 10^{-36}$ s. The value of $t_{\rm in}$ is insensitive to the variation of the initial conditions, provided that $t_{\rm in}$ is at least one order smaller than $t_{\rm en}$. With the adopted value of the e-folding $N$ and $\beta$, we can fix the values of $a_{\rm in}$ and of $t_{\rm in}$, respectively.
Since at the beginning of the inflationary stage the matter is not yet generated,  we have $\rho_{\rm in}= 0$. For the deceleration parameter, from Eq.~(\ref{q0}) we find
\begin{equation}
q\approx\frac{-\lambda_{0}^2e^{-2\beta a^2}-\lambda_1^2}{\lambda_{0}^{2}e^{-2\beta a^2}+\lambda_1^2}=-1.
\end{equation}
By substituting the tension we can rewrite Eq.~(\ref{3w}) as
\begin{equation}
\frac{d\rho}{dt }+3w\frac{1}{a}\frac{da}{dt }\rho=-\frac{d\lambda}{dt}=\frac{2\beta \lambda_0^2a\dot{a}e^{-2\beta a^2}}{\sqrt{\lambda_0^2e^{-2\beta a^2}-\frac{6{}^{(5)}\Lambda}{k_5^2}+\lambda_1^2}}.
\end{equation}
For $2\beta a^2\ll 1$, the exponential factor does not change much as $a$ increase. Therefore,
\begin{equation}
\frac{d\rho}{dt }+3w\frac{1}{a}\frac{da}{dt }\rho=\frac{2\beta \lambda_0^2a\dot{a}}{\sqrt{\lambda_0^2-\frac{6{}^{(5)}\Lambda}{k_5^2}+\lambda_1^2}}\approx \frac{2\beta \lambda_0^2a\dot{a}}{\sqrt{\lambda_0^2}}=\beta k_5^2\lambda_0^2 a^2/3.
\end{equation}
where we have also used Eq.~(\ref{approxa}). Conversion of the matter from the brane tension energy begins already during the inflationary stage, and the rate of the conversion is proportional to $a^2$ during inflation. Therefore, the matter density also rises exponentially in the late stages of the inflationary phase. The matter generation rate becomes most important at the end of inflation.

\subsection{Reheating period: $2\beta a^2 \approx 1$}

During the reheating period the evolution of the matter gradually changes from an exponential increase to a power law decrease, $\rho\propto a^{-3w}$. In our model the matter density is a smooth function, which is strictly increasing from the beginning of inflation, and then strictly decreases after the end of the reheating phase. Therefore there must be a maximum value of the density $\rho_{\rm max}$ at a time $t_{\rm max}$. After $t_{\rm max}$, the Universe was dominated by matter which is in the form of radiation, and almost all the energy of the brane tension converted into matter. The temperature of matter, corresponding to a radiation dominated Universe at $t_{\rm max}$, is denoted $T_{\rm RH}$, and is given by
\begin{equation}
\rho_{\rm max}=\frac{\pi^2}{15}(kT_{\rm RH})^4,
\label{rmax}
\end{equation}
where $k$ is Boltzmann's constant. Current theory on gravitinos production constraints $T_{\rm RH}$ to be $T_{\rm RH}< 10^9-10^{10}\;{\rm GeV}$ \cite{Ellis, Kawasaki,Moroi}. Therefore the maximum density of the Universe must satisfy the condition $\rho_{\rm max}<10^{36}-10^{40}\;{\rm GeV}^4$. The maximum density can be obtained from the condition $\dot{\rho}|_{t=t_{\rm max}}=0$, and, with the use of Eq.~(14) it is given as a solution of the equation
\begin{equation}\label{rhom}
3w\frac{1}{a}\frac{da}{dt}\rho_{\rm max}=-\frac{d\lambda}{dt }=\left.\frac{2\beta \lambda_0^2a\dot{a}e^{-2\beta a^2}}{\sqrt{\lambda_0^2e^{-2\beta a^2}-\frac{6{}^{(5)}\Lambda}{k_5^2}+\lambda_1^2}}\right|_{t=t_{\rm max}}.
\end{equation}
With the rough approximation $\beta a^2|_{t=t_{\rm max}}\approx \beta a_{\rm en}^2\backsim 1$, Eq.~(\ref{rhom}) can be written as
\begin{equation}
3w\rho_{\rm max}=\left.\frac{2\beta \lambda_0^2a^2e^{-2\beta a^2}}{\sqrt{\lambda_0^2e^{-2\beta a^2}-\frac{6{}^{(5)}\Lambda}{k_5^2}+\lambda_1^2}}\right\vert_{t=t_{\rm max}}\approx\left.\frac{2\lambda_0^2e^{-2\beta a^2}}{\sqrt{\lambda_0^2e^{-2\beta a^2}-\frac{6{}^{(5)}\Lambda}{k_5^2}+\lambda_1^2}}\right\vert_{t=t_{\rm max}}\approx 0.74\lambda_0.
\end{equation}
This relation gives the maximum matter density of the Universe. And the value of $\lambda_0\backsim 10^{39}{\rm GeV}^4$ that we have chosen is consistent with the maximum value of the matter energy density.

\subsection{Matter Domination period: $2\beta a^2 \gg 1$ and $2\lambda\rho \gg \lambda_1^2 $}

At this stage, the Universe is dominated by matter and the brane tension is roughly a constant. During this period, the key difference with the conventional cosmological models is the presence of the quadratic term in matter density, which will dominate the dynamics of the Universe at the beginning of this period. The evolution equation of the scale factor is
\begin{equation}
\frac{da}{dt}=a\frac{k_5^2}{6}\left(\sqrt{2\lambda\rho+\rho^2}\right)\approx a\frac{k_5^2}{6}\rho,
\label{dadt2}
\end{equation}
and the evolution of the matter density is given by
\begin{equation}
\frac{d\rho}{dt }+4\frac{1}{a}\frac{da}{dt }\rho=0,
\end{equation}
where we have assumed that immediately after the matter energy reaches its maximum the matter is in the form of radiation. The solution is $\rho={\rm constant} /a^4\equiv \rho_{\rm max}a^4(t_{\rm max})/a^4\approx \lambda_0/\beta^2a^4$. Therefore, the scale factor and deceleration parameter evolve as
\begin{equation}
a(t)=\left(\frac{2k_5^2}{3}\rho_{\rm max}a^4(t_{\rm max})t\right)^\frac{1}{4}\approx\frac{k_5^2\lambda_0}{6\beta^2}t,
\end{equation}
\begin{equation}
q=-\frac{a\ddot{a}}{\dot{a}^2}=\frac{\frac{k_5^4\lambda}{36}\left[\rho(1+\frac{2\rho}{\lambda})+3(w-1)\rho(1+\frac{\rho}{\lambda})\right]}{\frac{k_5^4\lambda}{18}(\rho+\frac{\rho^2}{\lambda})}\approx 1+3w=5.
\end{equation}
After the quadratic density and linear density equality $2\lambda\rho= \rho^2$, the linear term in matter will take over, i.e. $2\lambda\rho\gg \rho^2$. The Universe enters in the $\rm \Lambda CDM$ model at this radiation dominated phase, and its dynamics is described by the equations
\begin{equation}
\frac{da}{dt}=a\left(\sqrt{\frac{k_5^4}{18}\lambda\rho}\right),
\label{dadt3}
\end{equation}
and
\begin{equation}
\frac{d\rho}{dt }+4\frac{1}{a}\frac{da}{dt }\rho=0,
\end{equation}
respectively. The solution for the scale factor is
\begin{equation}
\frac{a^2(t)}{2}\approx\left(\frac{k_5^3\sqrt{-6{}^{(5)}\Lambda}\lambda_0}{18\beta^2}\right)^{1/2}t,
\label{alinear}
\end{equation}
with $\rho=\rho_{\rm max}a^4(t_{\rm max})/a^4$. During this period the deceleration parameter evolves as
\begin{equation}
q\approx\frac{\frac{k_5^4\lambda}{36}\left[\rho+3(w-1)\rho\right]}{\frac{k_5^4}{18}\lambda\rho}=\frac{3w-2}{2}=1.
\label{qlinear}
\end{equation}
After this stage, the Universe switches from radiation dominated to non-relativistic matter dominated at $t_{\rm rm}=1.5\times 10^{12}\;{\rm s}$ \cite{Ryden}. To simulate the transition from the radiation dominated period to the baryonic matter dominated period one can introduce a time varying $w$ given by \cite{Harko08}
\begin{equation}
w=\frac{4t_{\rm rm}/3+t}{t_{\rm rm}+t}.
\label{rmsim}
\end{equation}

In the matter dominated era ($w=1$), the deceleration parameter shifts to $q=0.5$, according to Eq.~(\ref{qlinear}). Recall that $\beta$ is still free, and we can use it to match the scale factor at the radiation-matter equality. With the use of Eq.~(\ref{alinear}), and by taking $a_{\rm rm}=2.8\times10^{-4}$ \cite{Ryden}, we obtain
\begin{equation}
2.6\times10^{-20}s^{-1}\backsim\left(\frac{k_5^3\sqrt{-6{}^{(5)}\Lambda}\lambda_0}{18\beta^2}\right)^{1/2},
\label{matchrm}
\end{equation}
which gives $\lambda_0\approx 10^{-15}\times \beta^2$. This gives the estimate of $\beta\backsim 10^{45}$.

\subsection{Dark Energy Domination era ($2\lambda\rho<\lambda_1^2 $)}

With the increase of the cosmological time, the constant term $\lambda_1$ in Eq.~(\ref{1}) will dominate over matter. Thus this term plays the role of the dark energy of the standard $\rm \Lambda CDM$ models. In its late stages of evolution the Universe becomes ``dark energy" dominated. The Universe turn to an exponential acceleration again, with
\begin{equation}
a\propto e^{(k_5^2/6)\lambda_1\, t},
\end{equation}
but with a time scale much longer than the inflationary time scale. The deceleration parameter will converge to
\begin{equation}
q\rightarrow\frac{-k_5^4\lambda_1^2}{k_5^4\lambda_1^2}=-1,
\end{equation}
showing that the expansion of the universe is accelerating.

\section{Numerical analysis of the model}\label{numerical}

The field equation can be rewritten in a simple form by introducing as set of dimensionless variables $\tau$, $l$, and $r$, defined as
\begin{equation}
\tau =k_{5}\sqrt{\frac{%
(-\Lambda _{5})}{2}}t,\rho =k_{5}^{-1}\sqrt{3\times (-\Lambda _{5})}%
r,\lambda =k_{5}^{-1}\sqrt{3\times (-\Lambda _{5})}l,  \label{dim}
\end{equation}
respectively. Then the field equations Eqs.~(\ref{1})-(\ref{5}) can be written in a dimensionless form as
\begin{equation}
3\left( \frac{1}{a}\frac{da}{d\tau }\right) ^{2}=-1+\frac{l^{2}}{2}+lr+\frac{%
r^{2}}{2},  \label{f1}
\end{equation}
\begin{equation}
\frac{1}{a}\frac{d^{2}a}{d\tau ^{2}}=-\frac{1}{3}+\frac{l^{2}}{6}-\frac{1}{6}%
l\left[ r\left( 1+\frac{2r}{l}\right) +3(w-1)r\left( 1+\frac{r}{l}\right) \right] ,  \label{f2}
\end{equation}
\begin{equation}
\frac{dr}{d\tau }+3w\frac{1}{a}\frac{da}{d\tau }r=-\frac{dl}{d\tau }.
\label{f3}
\end{equation}
By rescaling the four-dimensional cosmological constant so that $%
^{(4)}\Lambda =\left[ k_{5}^{2}(-\Lambda _{5})/2\right] l_{eff}$, it follows
that $l_{\rm eff}=-1+l^{2}/2$. In the dimensionless variables, the deceleration parameter is
given by $q=-\left( ad^{2}a/d\tau ^{2}\right) /\left( da/d\tau \right) ^{2}$%
, and can be explicitly expressed as a function of the physical parameters
of the model as
\begin{equation}
q=\frac{1-l^{2}/2+l\left[ r\left( 1+2r/l\right) +3(w-1)r\left( 1+r/l\right)
\right] /2}{-1+l^{2}/2+lr+r^{2}/2}.
\label{q}
\end{equation}
To obtain the numerical solution, one should also consider the rescaled form of $\lambda$,
\begin{equation}
l^{2}=2(l_{0}^{2}e^{-2\beta a^2}+1+l_1^2),
\label{exptension}
\end{equation}
where $l_{0}$ and $l_1$ are constants, with $l_0 \approx 10^{25} \gg l_1\approx 10^{-31}$. Eq.~(\ref{rmsim}) becomes
\begin{equation}
w=\frac{4\tau_{\rm rm}/3+\tau}{\tau_{\rm rm}+\tau},
\label{scalermsim}
\end{equation}
with $\tau_{\rm rm}=(4.8\times10^{-3}\;{\rm eV})t_{\rm rm}=10^{25}$. Note that the conversion in Eq.~(\ref{dim}) can be written out numerically
\begin{equation}
t=(208\;{\rm eV}^{-1})\tau=(1.4\times10^{-13}\;{\rm s})\tau,\rho =(2.0\times10^{50}\;{\rm eV})r,\lambda =(2.0\times10^{50}\;{\rm eV})l.  \label{dim2}
\end{equation}

\begin{figure}[th]
\centering
\includegraphics[width=.450\textwidth]{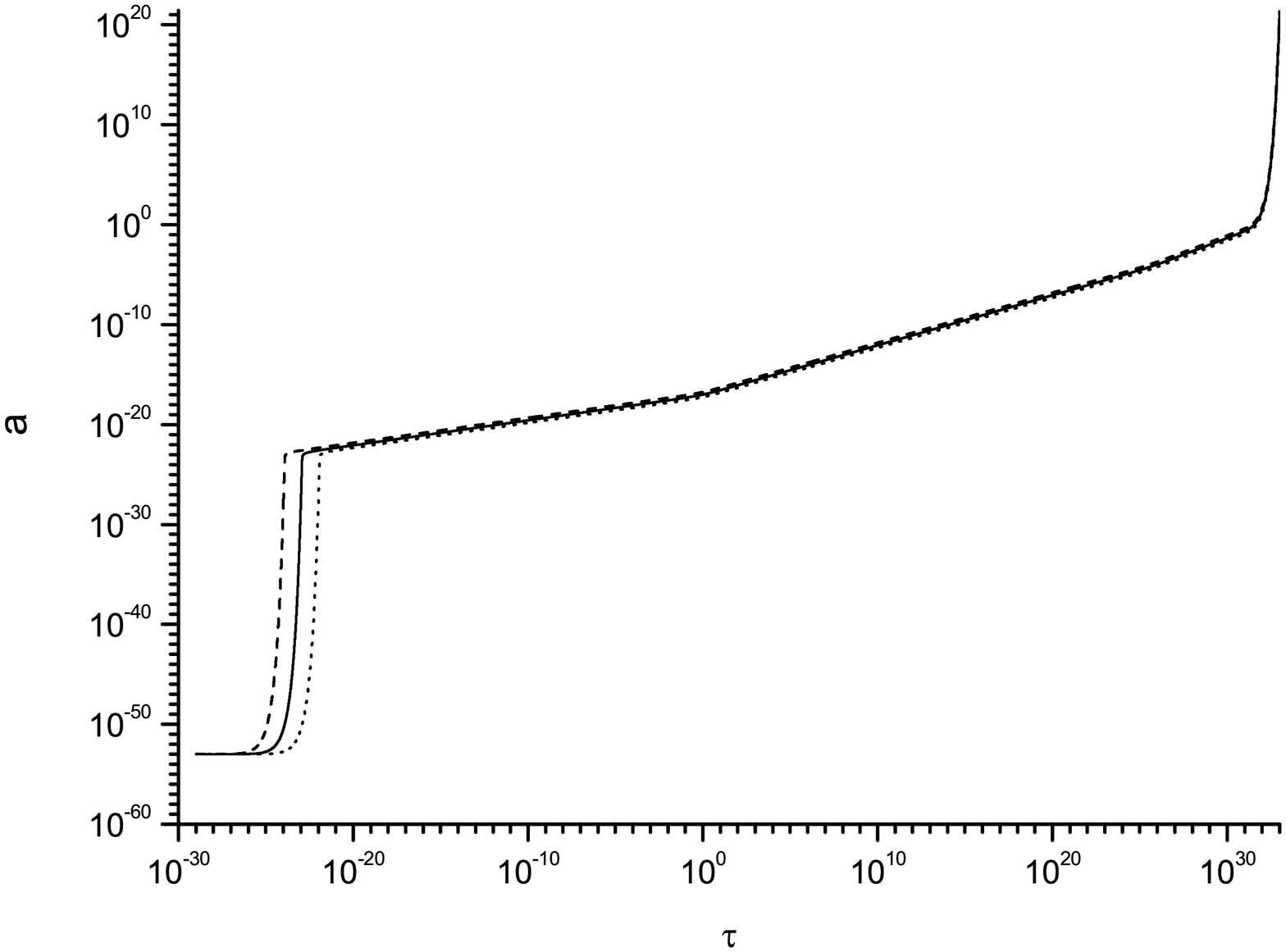}
\includegraphics[width=.450\textwidth]{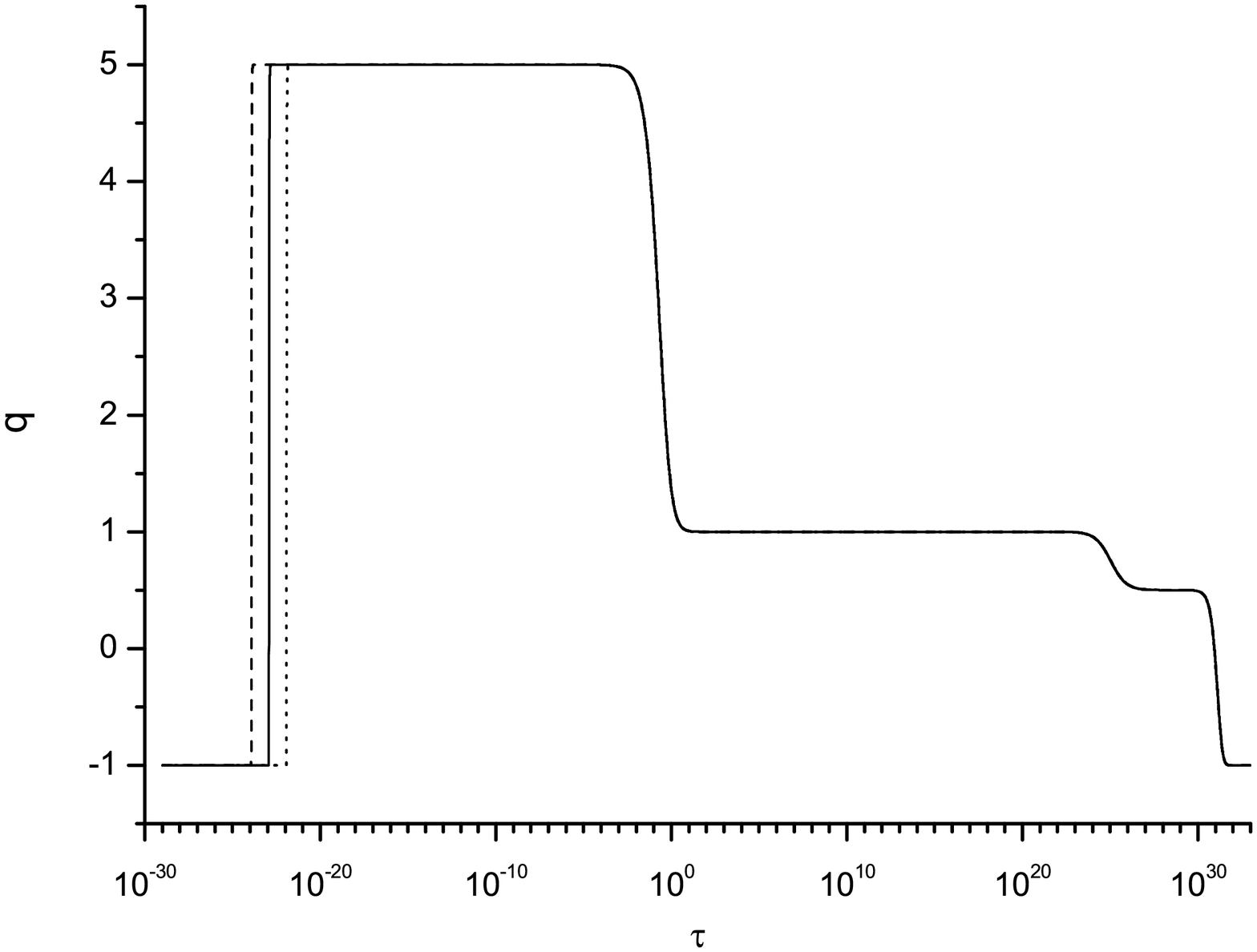}
\includegraphics[width=.450\textwidth]{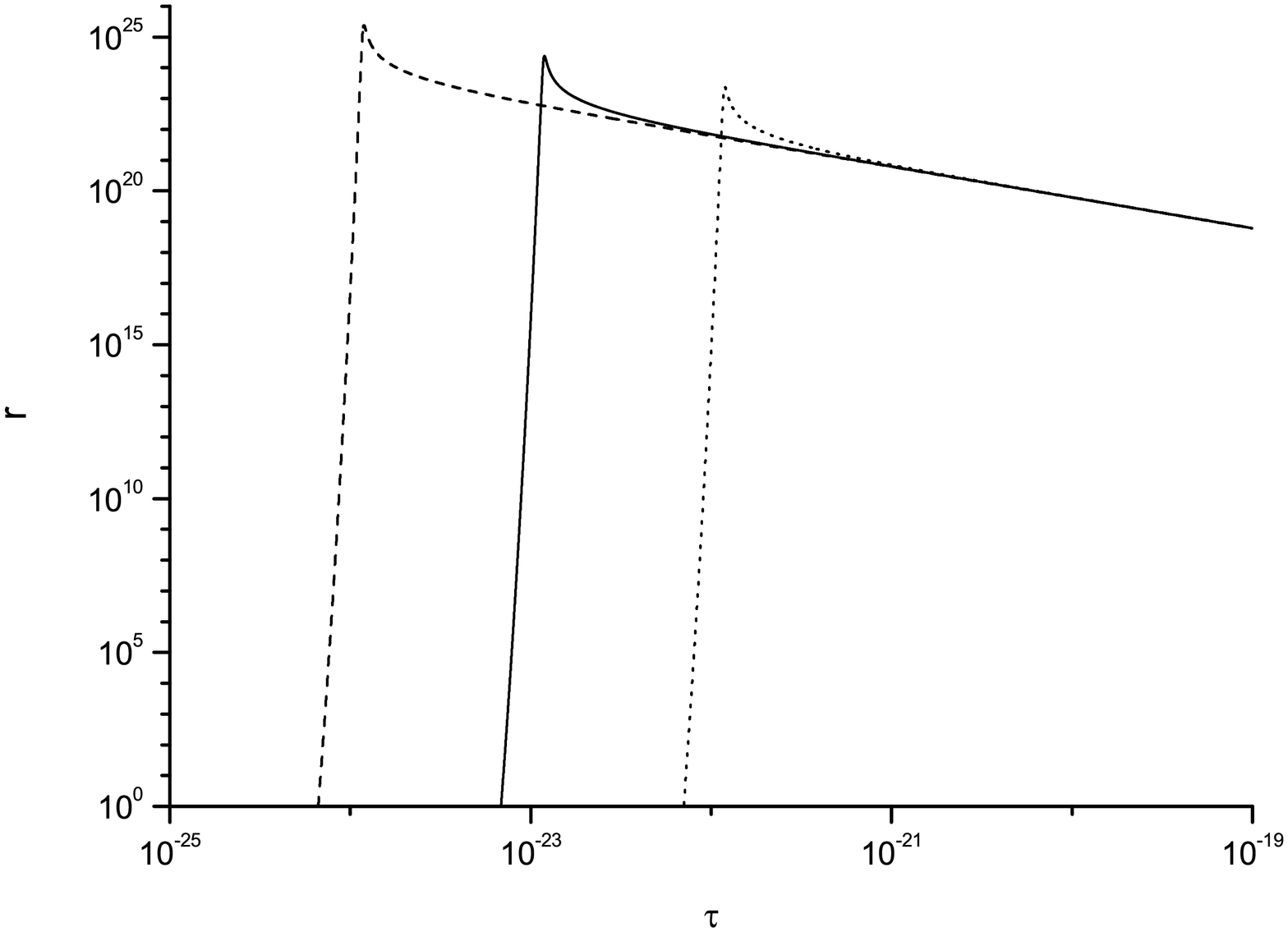}
\caption{Time variations of the scale factor $a$, deceleration parameter $q$, and energy density $r$ of the Universe in braneworld
models with varying scale factor dependent brane tension. Different $l_0$ are plotted: $l_0=10^{24}$ (dotted curve), $l_0=10^{25}$ (solid curve), and $l_0=10^{26}$ (dashed curve).}
\label{fig1}
\end{figure}

The time variations of the scale factor, of the energy density, of the brane
tension, and of the deceleration parameter of the Universe are presented,
for different scales of vacuum energy $l_0$, in Figs.~\ref{fig1}
respectively. From the plot of $a$ and $q$, we find that there are 5 stages in the evolution of the Universe. Namely, the inflation and reheating stage, the quadratic density stage, the radiation domination stage, the non-relativistic matter stage, and the late time acceleration stage. Comparing plots with different $l_0$'s, we see the effects of different vacuum energy scales on the evolution of the Universe. According to the plot with different $l_0$'s, we find that a larger vacuum energy would provide a faster inflation, and it also generates more matter. The matter energy density reaches its maximum at earlier times. Although there are many changes in the evolution of the Universe caused by changing $l_0$, these characteristics are hard to be constraint by observations.
On the other hand, $l_0$ cannot be arbitrary due to the theoretical constraint, e.g. vacuum energy density, gravitinos production, etc.
%We see that the evolution of $a$, $r$ and $q$ converge at a time right after inflation.

\begin{figure}[th]
\centering
\includegraphics[width=.450\textwidth]{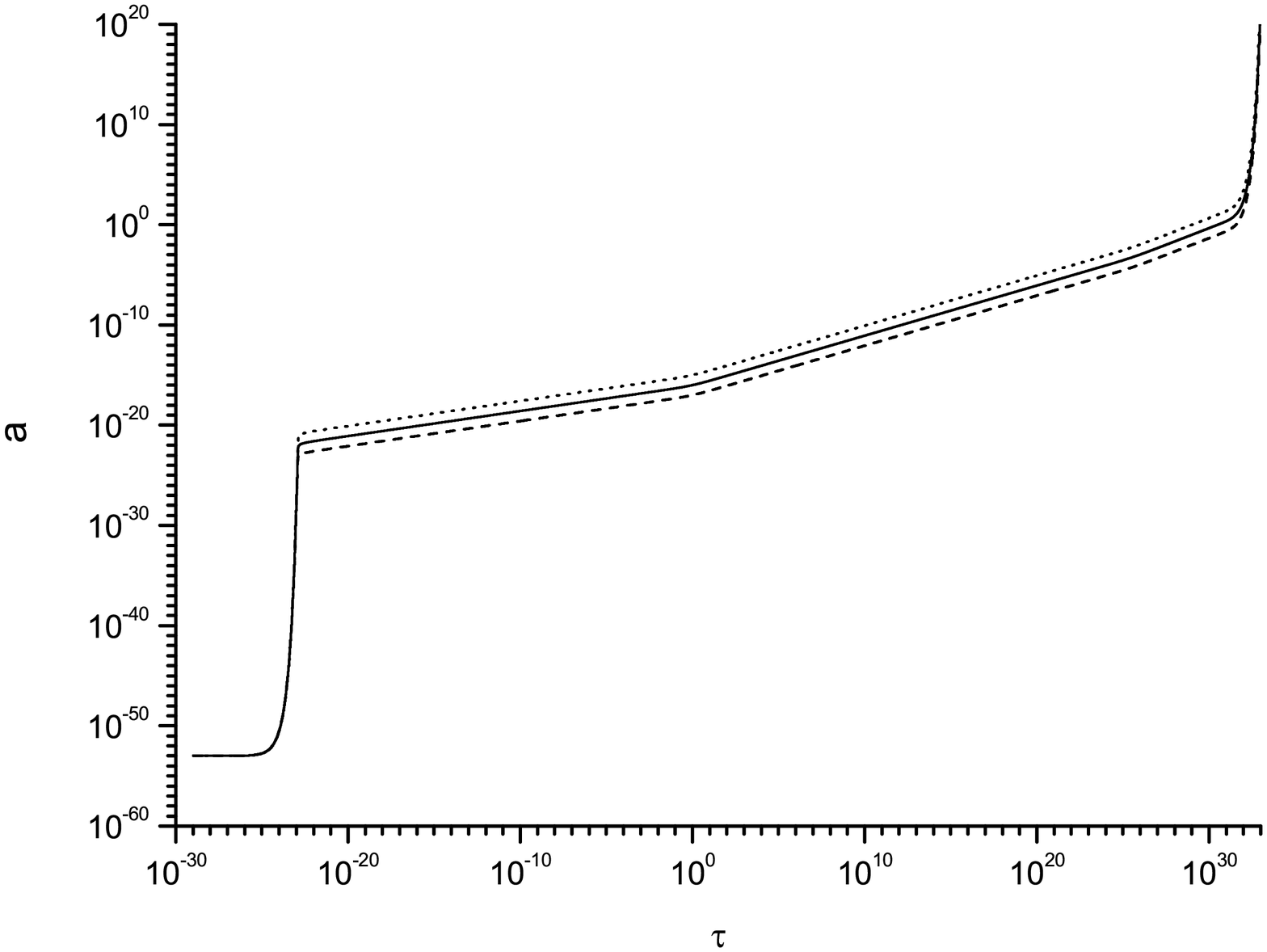}
\includegraphics[width=.450\textwidth]{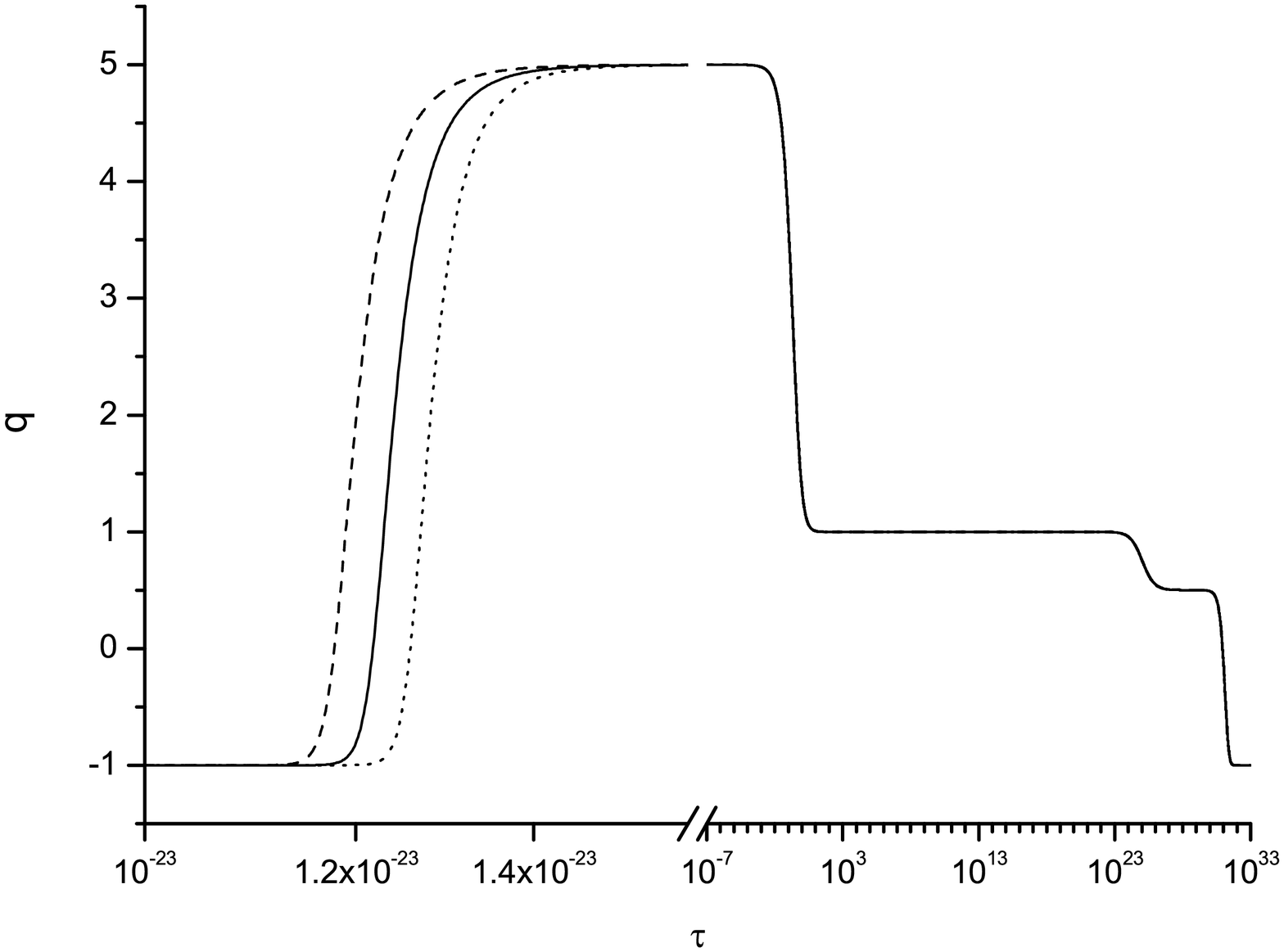}
\includegraphics[width=.450\textwidth]{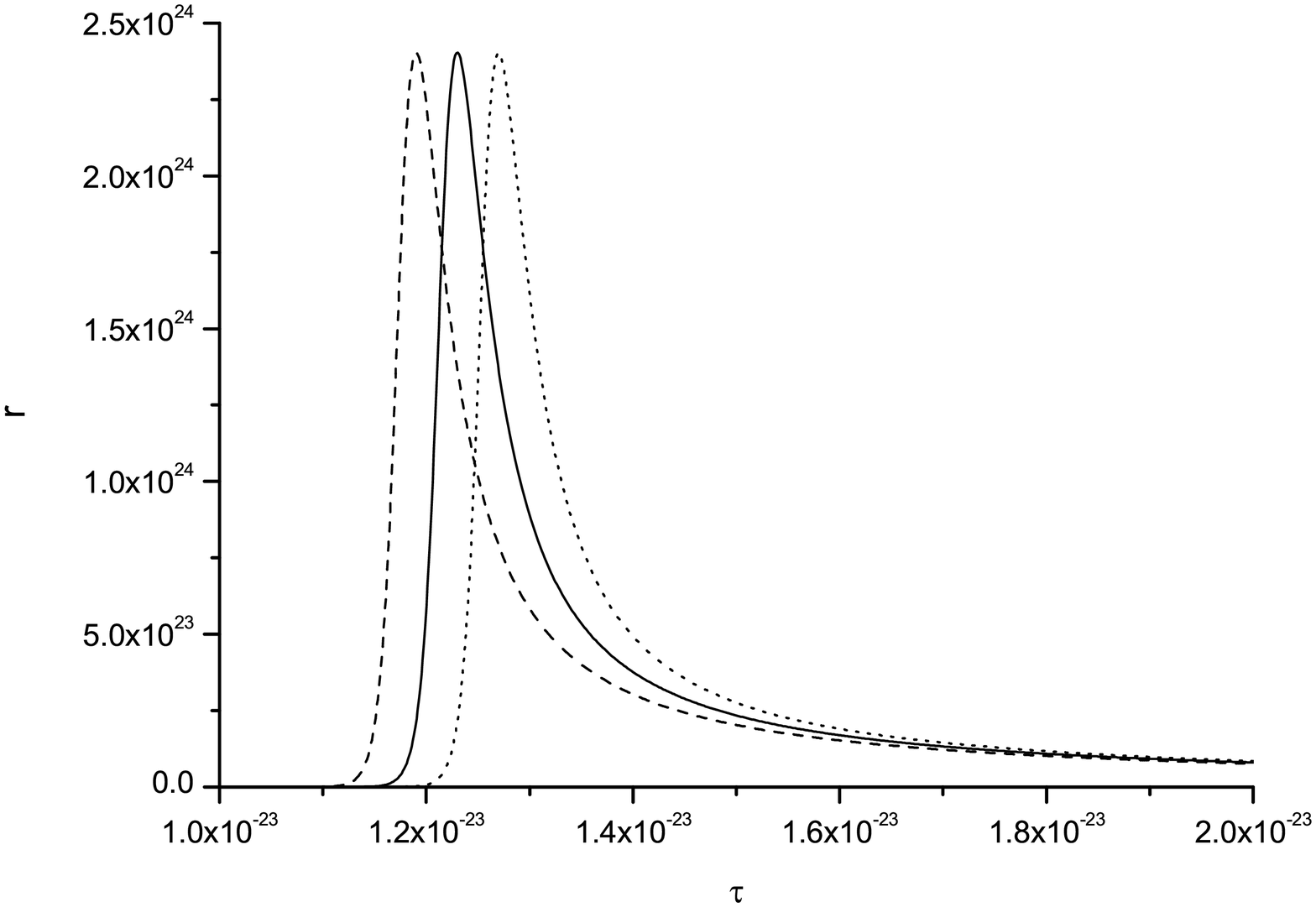}
\caption{Time variations of the scale factor $a$, deceleration parameter $q$, and energy density $r$ of the Universe in braneworld
models with varying scale factor dependent brane tension. Different $\beta$ are plotted: $\beta=10^{43}$ (dotted curve), $\beta=10^{45}$ (solid curve), and $\beta=10^{47}$ (dashed curve).}
\label{fig2}
\end{figure}

\begin{figure}[th]
\centering
\includegraphics[width=.63\textwidth]{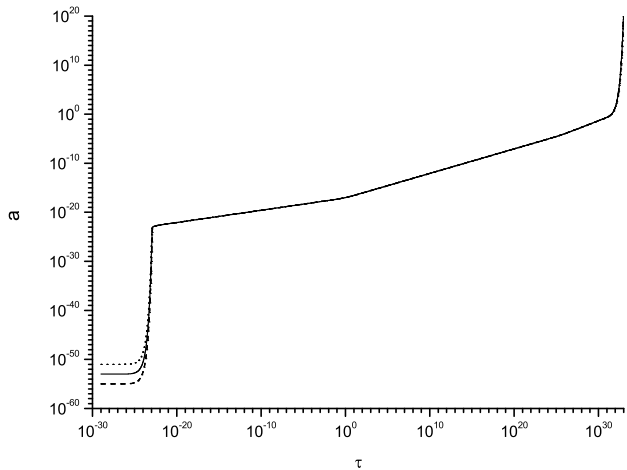}
\caption{Time variations of the scale factor $a$ of the Universe in braneworld
models with varying scale factor dependent brane tension. Different $a_{\rm in}$ are plotted: $a_{\rm in}=10^{-51}$ (dotted curve), $a_{\rm in}=10^{-53}$ (solid curve), and $a_{\rm in}=10^{-55}$ (dashed curve).}
\label{fig3}
\end{figure}

The time variations of the scale factor, and of the deceleration parameter of the Universe are presented,
for different values of $\beta$, in Figs.~\ref{fig2}
respectively. If we assume $\lambda_0$ is fixed by the vacuum energy scale, the value of $\beta$ is well confined. Varying $\beta$ would affect the observational constraint of $a$. From the graph of $r$, we can cross check that the value of $\rho_{\rm max}$ indeed fulfills the requirement of Eq.~(\ref{rmax}). Besides $\beta $, we could examine the effect of adopting a different e-folding $N$. Since $a_{\rm en}$ is determined by $\beta$ according to the condition Eq.~(\ref{aen}), it follows that it is determined by observational constraints. Different e-foldings could be a result of differences in $a_{\rm in}$. However, we find from Fig.~\ref{fig3} that $a_{\rm in}$ does not affect the post-inflationary epochs. Therefore $a_{\rm in}$ is not a robust parameters, and we cannot fully determine it.

\section{Discussions and final remarks}\label{conclusion}

In the present paper we have considered some cosmological implications of the braneworld models with variable tension. We have considered a thermodynamic interpretation of the model, and we have shown that from a thermodynamic point of view a variable brane tension can describe particle production processes in the early Universe, as well as the entropy production in the early stages of the cosmological evolution. A simple power-law inflationary cosmological model has also been obtained.   By adopting a simple analytical form for $\lambda$, we have obtained a complete description of the dynamics and evolution of the Universe from the stage of the inflation to the phase of the late acceleration.
Moreover, the differences between the energy scales of the theoretical vacuum energy at inflationary epoch and during late time acceleration are studied.

A variable brane tension can drive the inflationary evolution of the Universe, and also be responsible for matter creation in the post-inflationary phase (the reheating of the Universe). In the model we have adopted the brane tension converges to a constant that gives the gravitational constant, and the residue after cancelation with the $\Lambda_5$ term would give the dark energy. It is interesting to note that, as opposed to the standard cosmological scenarios, in the present model matter creation takes place during the entire inflationary phase, but reaches a maximum after the exponential expansion of the Universe ends. Therefore in the braneworld models with varying brane tension there can be no clear distinction between inflation and reheating. By using numerical analysis as well as approximate analytical methods, we have obtained the result that in this variable tension model there are 5 phases in the cosmological evolution of the Universe. At the beginning, there are the inflationary and the reheating phases. During these phases, the brane tension is the dominant energy in the Universe, and its magnitude is of the order of the vacuum energy at GUT scale. As the brane tension decays, the matter is created, and the Universe enters in the hot Universe phase of the Big Bang picture. The third phase is the quadratic density domination phase. This is a unique characteristic of the brane world scenario, during which the Universe is dominated by a quadratic term in the energy density of the radiation. The deceleration parameter increases from $-1$ during inflation to $5$ at this phase.
To study the cosmological dynamics we have introduced a set of dimensionless quantities, which describe the evolution of the scaled densities in terms of a dimensionless time parameter $\tau$. By using the results of the numerical simulations for the rescaled variables, one can obtain some constraints on the physical parameters of the model.

\section*{Acknowledgments}

This work is supported by the RGC grant HKU 701808P of the Government of the
Hong Kong SAR.


\begin{thebibliography}{99}


\bibitem{Randall}  L. Randall and R. Sundrum, Phys. Rev. Lett \textbf{83},
3370 (1999).

\bibitem{Randall2}  L. Randall and R. Sundrum, Phys. Rev. Lett \textbf{83},
4690 (1999).

\bibitem{Witten}  P. Horava and E. Witten, Nucl. Phys. \textbf{B460}, 506
(1996).

\bibitem{Witten2}  P. Horava and E. Witten, Nucl. Phys. \textbf{B475}, 94
(1996).

\bibitem{Polchinski}  J. Polchinski, Phys. Rev. Lett \textbf{75}, 4724
(1995);

\bibitem{Shiromizu}  M. Sasaki, T. Shiromizu and K. Maeda, Phys. Rev.
\textbf{D62}, 024008 (2000); T. Shiromizu, K. Maeda and M. Sasaki, Phys.
Rev. \textbf{D62}, 024012 (2000); K. Maeda, S. Mizuno and T. Torii, Phys.
Rev. \textbf{D68}, 024033 (2003).

\bibitem{all2}  P. Bin\'{e}truy, C. Deffayet and D. Langlois, Nucl. Phys. B
\textbf{565}, 269 (2000); R. Maartens, Phys. Rev. \textbf{D62}, 084023
(2000); A. Campos and C. F. Sopuerta, Phys. Rev. \textbf{D63}, 104012
(2001); A. Campos and C. F. Sopuerta, Phys. Rev. \textbf{D64}, 104011
(2001); C.-M. Chen, T. Harko and M. K. Mak, Phys. Rev. \textbf{D64}, 044013
(2001); D. Langlois, Phys. Rev. Lett. \textbf{86}, 2212 (2001); C.-M. Chen,
T. Harko and M. K. Mak, Phys. Rev. \textbf{D64}, 124017 (2001); J. D. Barrow
and R. Maartens, Phys. Lett. \textbf{B532}, 153 (2002); C.-M. Chen, T.
Harko, W. F. Kao and M. K. Mak, Nucl. Phys. \textbf{B636}, 159 (2002); M.
Szydlowski, M. P. Dabrowski and A. Krawiec, Phys. Rev. \textbf{D66}, 064003
(2002); T. Harko and M. K. Mak, Class. Quantum Grav. \textbf{20}, 407
(2003); C.-M. Chen, T. Harko, W. F. Kao and M. K. Mak, JCAP \textbf{0311},
005 (2003); T. Harko and M. K. Mak, Class. Quantum Grav. \textbf{21}, 1489
(2004); M. K. Mak and T. Harko, \prd {\bf 70}, 024010 (2004); T. Harko and
M. K. Mak, Phys. Rev. \textbf{D69}, 064020 (2004); A. N. Aliev and A. E.
Gumrukcuoglu, Class. Quant. Grav. \textbf{21}, 5081 (2004); M. Maziashvili,
Phys. Lett. \textbf{B627}, 197 (2005); S. Mukohyama, Phys. Rev. \textbf{D72}%
, 061901 (2005); M. K. Mak and T. Harko, Phys. Rev. \textbf{D71}, 104022
(2005); L. A. Gergely and Z. Kovacs, Phys. Rev. {\bf D72}, 064015 (2005); A. N. Aliev and A. E. Gumrukcuoglu, Phys. Rev. \textbf{D71}, 104027 (2005); T. Harko and K. S. Cheng, Astrophys. J. \textbf{636}, 8 (2006); L.
A. Gergely, Phys. Rev. \textbf{D74} 024002, (2006); N. Pires, Zong-Hong Zhu,
J. S. Alcaniz, Phys. Rev. \textbf{D73}, 123530 (2006); C. G. B\"ohmer and T.
Harko, Class. Quantum Grav. \textbf{24}, 3191 (2007); M. Heydari-Fard and H.
R. Sepangi, Phys. Lett. \textbf{B649}, 1 (2007); T. Harko and K. S. Cheng,
Phys. Rev. \textbf{D76}, 044013 (2007); A. Viznyuk and Y. Shtanov, Phys.
Rev. \textbf{D76}, 064009 (2007); Z. Kovacs and L. A. Gergely, Phys. Rev.
\textbf{D77}, 024003 (2008); T. Harko and V. S. Sabau, Phys. Rev. \textbf{D77%
}, 104009 (2008); L. P. Chimento, M. Forte, and M. G. Richarte, Phys. Rev. {\bf D79}, 083527 (2009); V. G. Czinner and A. Flachi, Phys. Rev. {\bf D80}, 104017 (2009); I. Gurwich, S. Rubin, and A. Davidson, Phys. Lett. {\bf B679}, 515 (2009); N. E. Mavromatos, S. Sarkar, and W. Tarantino, Phys. Rev. {\bf D80}, 084046 (2009), Z. Keresztes and L. A. Gergely, Ann. Physik {\bf 19}, 249 (2010); Z. Keresztes and L. A. Gergely, Class. Quant. Grav. {\bf 27}, 105009 (2010).

\bibitem{Carroll} S. M. Carroll, Living Rev. Relativity {\bf 3}, 1 (2001).

\bibitem{Maartens} R. Maartens, Living Rev. Relativity {\bf 7}, 1 (2004).

\bibitem{Spergel}  D. N. Spergel \emph{et al.}, Astrophys. J. Supplement
Series \textbf{170}, 377 (2007).

\bibitem{Guth}  A. H. Guth, Phys. Rev. \textbf{D71}, 347 (1981).

\bibitem{infl}  A. Linde, Phys. Repts. \textbf{333-334}, 575 (2000); B. A.
Bassett, S. Tsujikawa and D. Wands, Rev. Mod. Phys. \textbf{78}, 537 (2006).

\bibitem{reh}  A. D. Dolgov and A. D. Linde, Phys. Lett. \textbf{B116}, 329
(1982); L. F. Abbott, E. Farhi and M. B. Wise, Phys. Lett. \textbf{B117}, 29
(1982); A. Albrecht, P. J. Steinhardt, M. S. Turner and F. Wilczek, Phys.
Rev. Lett. \textbf{48}, 1437 (1982).

\bibitem{Susperregi}  M. Susperregi, Phys. Rev. \textbf{D68}, 123509 (2003).

\bibitem{Liddle3}  A. R. Liddle and L. A. Ure\~{n}a-L\'{o}pez, Phys. Rev.
Lett. \textbf{97}, 161301 (2006).

\bibitem{Cardenas}  V. H. C\'{a}rdenas, Phys. Rev. \textbf{D75}, 083512
(2007).

\bibitem{Kolb}  E. W. Kolb, A. Notari and A. Riotto, Phys. Rev. \textbf{D68}%
, 123505 (2003).

\bibitem{HiTa03}  Y. Himemoto and T. Tanaka, Phys. Rev. \textbf{D67}, 084014
(2003).

\bibitem{BrDa03}  J. H. Brodie and D. A. Easson, JCAP \textbf{0312}, 004
(2003).

\bibitem{SaDaSh03}  M. Sami, N. Dadhich and T. Shiromizu, Phys. Lett.
\textbf{B568}, 118 (2003).

\bibitem{TaMa04}  Y. I. Takamizu and K. I. Maeda, Phys. Rev. \textbf{D70},
123514 (2004).

\bibitem{Hannestad}  S. Hannestad, Phys. Rev. \textbf{D70}, 043506 (2004).

\bibitem{Pa05}  G. Panotopoulos, JCAP \textbf{0508}, 005 (2005).

\bibitem{DaKh06}  E. Abou El Dahab and S. Khalil, JHEP \textbf{0609}, 042
(2006).

\bibitem{PaZa06}  E. Papantonopoulos and V. Zamarias, JCAP \textbf{0611},
005 (2006).

%\bibitem{Umezu}  K. Umezu, K. Ichiki, T. Kajino, G. J. Mathews, R. Nakamura
%and M. Yahiro, Phys. Rev. \textbf{D73}, 063527 (2006).

%\bibitem{BoTa06}  M. R. Setare, Phys. Lett. \textbf{B642}, 421 (2006); C.
%Bogdanos and K. Tamvakis, Phys. Lett. \textbf{B646}, 39 (2007).

%\bibitem{Bogdanos}  C. Bogdanos, S. Nesseris, L. Perivolaropoulos and K.
%Tamvakis, astro-ph 0705318.

\bibitem{Ger1}  L. \'A. Gergely, Phys. Rev. \textbf{D78}, 084006 (2008).

\bibitem{Ger2}  L. \'A. Gergely, Phys. Rev. {\bf D79}, 086007 (2009).
%\bibitem{Albrecht}
%A. Albrecht, P. J. Steinhardt, M. S. Turner and F. Wilczek, Phys. Rev. Lett. {\bf 48}, 1437 (1982).

\bibitem{yun} L. Barosi, F. A. Brito, and A. R. Queiroz, JHEP \text{0904}, 030 (2009); S. Yun, Mod. Phys. Lett. {\bf A25}, 159 (2010); M. Rogatko and A. Szyplowska, Gen. Rel. Grav. {\bf 42}, 209 (2010).

%\bibitem{Harko}  T. Harko and M. K. Mak, Astrophys. and Space Science,
%\textbf{253}, 161 (1997).

\bibitem{Harko08} T. Harko, W. F. Choi, K. C. Wong and K. S. Cheng, JCAP \textbf{0806}, 002(2008)

\bibitem{Ellis}  J. R. Ellis, J. E. Kim and D. V. Nanopoulos, Phys. Lett.
\textbf{B145}, 181 (1984).

\bibitem{Kawasaki}  M. Kawasaki and T. Moroi, Prog. Theor. Phys. \textbf{93}%
, 879 (1995).

\bibitem{Moroi}  M. Kawasaki and T. Moroi, Phys. Lett. {\bf B346},  27 (1995).

\bibitem{Ishak} M. Ishak, Month. Not. R. Astron. Soc. {\bf 363}, 469 (2005).


\bibitem{Ryden} B. Ryden, \emph{Introduction to Cosmology}, Addison Wesley, USA (2003).

\end{thebibliography}
\end{document}